\documentclass{article}

\usepackage{graphicx}
\usepackage{amsmath}
\usepackage{amssymb}
\usepackage{float}
\usepackage{placeins}
\usepackage{xcolor}
\usepackage{authblk}

\definecolor{visprocessgray}{RGB}{127,127,127}

\definecolor{RGcolor}{RGB}{49,121,198}

\definecolor{newsleakunitcolor}{RGB}{144,144,144}

\definecolor{frontbackendcolor}{RGB}{80,159,198}

\definecolor{dataTypeColor}{RGB}{150,98,208}

\definecolor{inputoutputcolor}{RGB}{22,166,3}

\definecolor{apicolor}{RGB}{148,23,81}

\definecolor{stepcolor}{RGB}{47,82,143}

\definecolor{lisaRed}{RGB}{237,28,35}

\definecolor{lisaBlue}{RGB}{62,70,200}

\definecolor{lisaBlue2}{RGB}{1,162,232}

\definecolor{lisaYellow}{RGB}{255,201,14}

\definecolor{lisaGreen}{RGB}{33,177,76}

\definecolor{colorbrewerorange}{RGB}{253,140,90}
\definecolor{colorbreweryellow}{RGB}{255,192,1}
\definecolor{colorbrewergreen}{RGB}{146,207,94}

\definecolor{C1}{RGB}{231,69,51}
\definecolor{C2}{RGB}{26,198,83}
\definecolor{C3}{RGB}{56,110,165}
\definecolor{C4}{RGB}{11,84,1}
\definecolor{C5}{RGB}{253,148,7}
\definecolor{C6}{RGB}{0,0,109}
\definecolor{C7}{RGB}{230,28,121}
\definecolor{C8}{RGB}{118,13,21}

\definecolor{darkYellow}{RGB}{255,210,0}

\definecolor{bc1}{HTML}{66C2A5}
\definecolor{bc2}{HTML}{FC8D62}
\definecolor{bc3}{HTML}{8DA0CB}

\definecolor{transcriptblack}{RGB}{0,0,0}

\definecolor{grayText}{RGB}{102,102,102}

\definecolor{factorOne}{RGB}{255,47,146}
\definecolor{factorTwo}{RGB}{3,176,240}

\definecolor{darkblue}{HTML}{005CBB}
\definecolor{darkgreen}{HTML}{307F25}
\definecolorseries{gradient}{rgb}{last}{black!3}{black!50}
\resetcolorseries[100]{gradient}

\usepackage{color}
\usepackage{booktabs}
\usepackage{calc}
\usepackage{colortbl}
\usepackage{comment}
\usepackage{ifthen}

\newcommand{\mybarecce}[1]{%
	\textcolor{black}{\rule{0.25pt+1pt*\real{#1}}{1.5ex}}\hfill
    \ifthenelse{\equal{#1}{0}}{\textcolor{black}{{\scriptsize #1}}}{\textcolor{black}{{\footnotesize #1}}}
}

\definecolor{cRanking}{HTML}{065700}

\newcolumntype{a}{>{\columncolor{black!10}}p{6.4em}}
\newcolumntype{b}{>{\columncolor{black!0}}p{6.4em}}

\definecolor{orange}{RGB}{255,127,0}

\definecolor{blue}{RGB}{0,128,255}

\definecolor{lilac}{RGB}{158,188,218}

\definecolor{myRed}{RGB}{177,36,24}

\definecolor{myLightBlue}{RGB}{75,174,234}

\definecolor{myLilac}{RGB}{191,162,209}

\definecolor{myGreen}{RGB}{76,173,91}

\newcommand{\replicatingExaminer}[1]{\emph{replicating examiner}\,}
\newcommand{\originalExaminer}[1]{\emph{original examiner}\,}
\newcommand{\originalExperiment}[1]{\emph{original experiment}\,}
\newcommand{\replicatingExperiment}[1]{\emph{replicating experiment}\,}

\definecolor{greenSC}{RGB}{0,176,80}

\definecolor{lilac2}{RGB}{104,52,154}

\definecolor{blue2}{RGB}{56,84,146}

\newcommand{\sampleselector}[1]{\textsc{{\large S}ample{\Large S}elector}}
\newcommand{\iv}[1]{independent variable}
\newcommand{\gv}[1]{grouping variable}
\newcommand{\cv}[1]{confounding variable}
\newcommand{\con}[1]{constraint}
\newcommand{\gss}[1]{goal sample size}
\newcommand{\es}[1]{experiment sample}
\newcommand{\evs}[1]{experiment variables}
\newcommand{\dpdh}[1]{data-property-driven hypothesis}
\newcommand{\erps}[1]{experiment-relevant properties}

\definecolor{orangeAlena}{RGB}{237,125,49}

\usepackage{amsmath}
\usepackage{amssymb}
\usepackage{color}
\usepackage[normalem]{ulem}
\usepackage{booktabs}
\usepackage{calc}
\usepackage{overpic}
\usepackage{colortbl}
\usepackage{hyperref}
\usepackage{gensymb}

\usepackage[anythingbreaks,hyphenbreaks]{breakurl}
\usepackage{makecell}

\extrafloats{100}

\usepackage{subfig}

\usepackage{enumitem}
\definecolor{yellow_min_cirlcle}{RGB}{255,192,0}

\definecolor{blue_max_cirlcle}{RGB}{0,176,240}

\definecolor{red_large_cirlcle}{RGB}{255,0,0}

\begin{document}
\title{GuidelineExplorer -- Navigating through the Forrest of Actionable Guidelines on Node-Link Graph Visualization}

\author[1]{Kathrin Guckes (n\'{e}e Ballweg)}{Kathrin.Ballweg@gris.tu-darmstadt.de}
\author[1]{Lisa Eisenhardt}{lisa\_eisenhardt@yahoo.de}
\author[2]{Prof. Margit Pohl}{margit.pohl@tuwien.ac.at}
\author[3]{Prof. Tatiana von Landesberger}{landesberger@cs.uni-koeln.de}

\affil[1]{Graphical Interactive Systems Group, Technical University Darmstadt}

\affil[2]{Informatics, TU Wien}


\affil[3]{Visualisierung und Visual Analytics, University of Cologne}


\maketitle

\begin{abstract}
Creating graph visualizations involves many decisions, such as layout, node and edge appearance, and color choices. These decisions are challenging due to the multitude of options available. For instance, graph layout can be force-directed or orthogonal, and edges can be curved, tapered, partially drawn, or animated. Thus, research offers a multitude of guidelines to optimize graph visualizations for human perception and usability. Guidelines can be actionable, providing direct instructions, or non-actionable, specifying what to avoid. This work focuses on actionable guidelines for node-link diagrams, aiding designers in making better decisions. 

Given the abundance of graph visualization research and the difficulty in navigating it, this work aims to collect and structure actionable guidelines for node-linkvisualizations. To demonstrate the general applicability of our approach to structuring actionable guidelines for node-link diagrams, we also included guidelines for visualizing graphs as matrices. It also proposes a visual interactive system, GuidelineExplorer, to apply guidelines directly to graphs, streamlining the design process and promoting collaboration within the research community.
\end{abstract}

  \section{Introduction} 
\label{sec:motivation_and_problem_statement}
Many decisions are necessary to create a graph visualization. For example, decisions about the layout of the graph, the appearance of the nodes and edges or whether and with which colors the nodes should be displayed. Making such decisions is not easy because there is a manifold option pool for all these decision areas. For example, the layout of the graph can be implemented as a force-directed layout \cite{Pohl09} or as an orthogonal layout \cite{Kieffer2016}. The edges in a node-link diagram can also be displayed in different ways; for example, curved \cite{10.1007/978-3-642-31223-6_34}, tapered \cite{10.1145/1518701.1519054}, only partially drawn \cite{10.1007/978-3-642-25878-7_22} or as an animated pattern \cite{5742390}.

\begin{figure}[tb]
	\centering
	\includegraphics[width=0.7\textwidth]{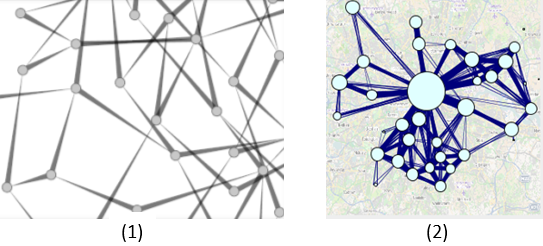}
	\caption[The effect of data properties on postulated actionable guidelines]{The effect of data properties on postulated actionable guidelines -- the tapered edges from \cite{10.1145/1518701.1519054} (1) do not perform well for the geo-located mobility graph of \cite{Landesberger16} (2).}
	\label{fig:Landesberger16-konischeKanten}
\end{figure}

Thus, there is a lot of research on how graphs can be visualized best so humans can perceive and work with them as well as possible. These results are published as guidelines. Guidelines can be divided into actionable and non-actionable guidelines. While actionable guidelines give the visualization designer direct instructions for action, non-actionable guidelines only specify what the visualization designer should refrain from doing in order not to provoke perceptual or cognitive resp. interpretative problems. She still does not know -- in the case of non-actionable guidelines -- what to do to make the visualization well perceivable and good to work with. The focus of this work is on actionable guidelines for node-link diagrams, since they give direct instructions and thus help the visualization designer to make better design decisions. In order to show that our approach of structuring actionable guidelines for node-link diagrams is generalizable, we added guidelines dealing with the visualization of graphs as matrices.

The visualized graphs, on which the guidelines are based, are created from datasets that have certain properties; e.g., the number of nodes or the graph density. These properties have an influence on how well a guideline performs or not. Figure \ref{fig:Landesberger16-konischeKanten} shows an example of the influence of data properties on the functional quality of guidelines. Holten et al. \cite{10.1145/1518701.1519054} have found that tapered edges improve the readability visualized direction of directed edges. Therefore von Landesberger et al. wanted to follow this guideline and use it for their mobility graphs. However, they found that the recommended tapered edges do not work due to extensive overplotting for their geo-located mobility graphs (cf. Figure~\ref{fig:Landesberger16-konischeKanten} -- \texttt{(2)}). The graphs Holten et al. used to postulate the guideline were not geo-located, directed graphs. Thus, it was possible for the authors to use a layout algorithm that freely chooses the position of the nodes and thus avoids overplotting. In addition, the graphs used by Holten et al. were also more sparse than von Landesberger et al.'s mobility graphs.
\FloatBarrier

In the field of graph visualizations there is a lot of research that formulates guidelines. There are so many that it is difficult to get an overview of the research and to find out where it is still worthwhile to continue research. It is also difficult for visualization designers to find suitable guidelines that support them in their design decisions. Therefore the goal of this work is to collect and structure actionable guidelines. A further goal is to implement a visual interactive system where guidelines can be applied directly to a graph. This allows the designer of a visualization to see directly how her own graph looks like with the guideline's recommendation without any implementation effort. Such a system would have saved Landesberger et al. \cite{Landesberger16} to implement the tapered edges guideline themselves. With our implementation -- GuidelineExplorer -- we suggest that such a system could become a community effort by having the authors of a guideline formulating paper implement their guideline for the system, thus saving their colleagues effort and time in the initial testing phase of the visual mappings. 

\section{Taxonomical Perspectives on Actionable Guidelines} 
\label{sec:taxonomicalPerspectives}
Our approach is to classify the guidelines into a taxonomy. We have different perspectives within our taxonomy. Those perspective help the visualization designer to choose a suitable guideline and thus make an appropriate design decision based on the information or question she currently has at hand. This might be ``How to appropriately design a specific part of a visualization?'' (cf. Section~\ref{sub:basic_taxonomy} -- \texttt{Foundational Perspective}). For visualizing a directed graph with a node-link diagram this might be ``How to visualize the direction of the edges best?'' as normal arrow-based edges can produce a considerable amount of clutter around the nodes so that the direction readability suffers (cf. Section~\ref{sub:graph_type_perspective}). It might also be ``How to design the visualization so that it properly supports a specific task?'' (cf. Section~\ref{sub:task_perspective} -- \texttt{Task Perspective}) or ``How to design a visualization so that users can quickly come to an answer for analyses under time pressure?'' (cf. Section~\ref{sub:if_type_perspective} -- \texttt{If-Type Perspective}).\\
As the foundational taxonomical perspective, we have chosen to structure the guidelines according to the the visualization decisions they deal with. 
In our opinion, this choice of a foundational taxonomy perspective has the following advantage: Here, the focus is on the necessary decisions to be made. 
And -- ``When does a visualization designer consult guidelines?'' -- commonly when she has to make a certain visualization decision for a certain part of her visualization. It can be, for instance for the large research areas of dynamic graphs or graph aesthetics, that there are not all visualization decisions present yet. This, however, is not an issue. Our taxonomical perspective provides the  bottom-up created framework which allows the addition of (sub-)categories as it was created with a thematic analysis \cite{doi:10.1191/1478088706qp063oa,thematicAnalysis}.


%
%
%
%
%
%
%

\subsection{Visualization Decision -- Foundational Perspective} 
\label{sub:basic_taxonomy}

\begin{figure}[tb]
  \centering
    \includegraphics[width=.9\textwidth]{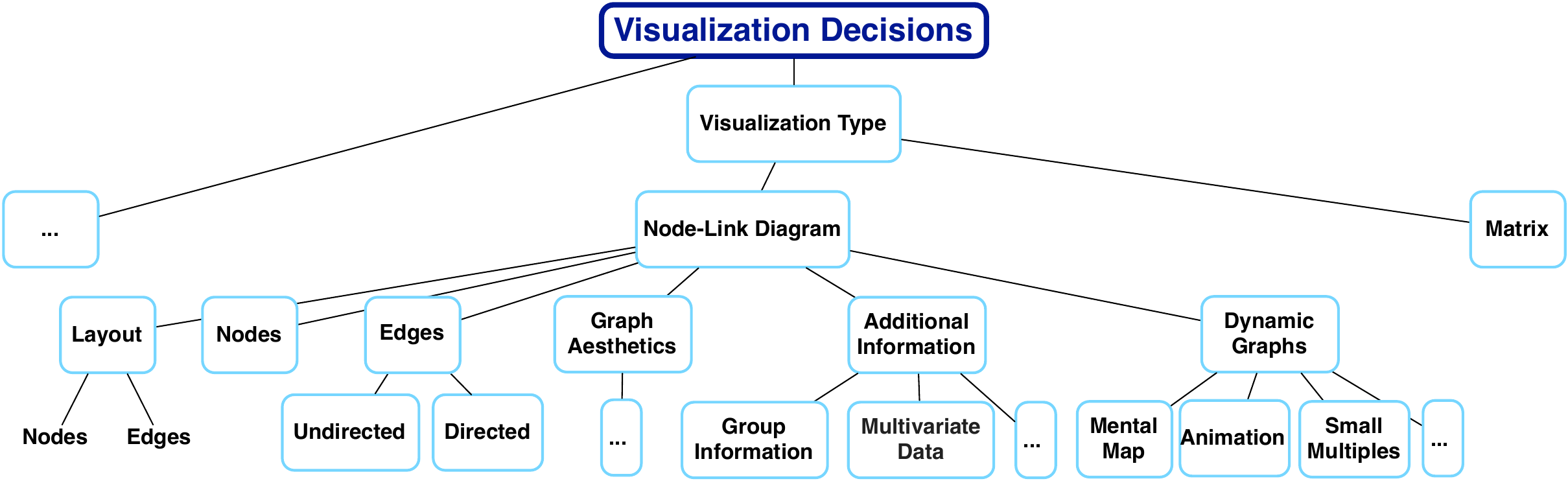}
  \caption{Visualization decision -- foundational taxonomy perspective}
  \label{fig:pictures_DACH_Chapter3_Perspective_VisualizationDecision}
\end{figure}

Figure~\ref{fig:pictures_DACH_Chapter3_Perspective_VisualizationDecision} shows the structure of our foundational perspective. As stated in the introduction, the focus of our work are actionable guidelines on node-link diagrams. The fact that we were able to incorporate the guidelines for matrices not only in our foundational taxonomy perspective but also in the alternative perspectives shows the generalizability of our taxonomical perspectives across visualization types.

A node-link diagram always has a \texttt{layout} that determines how the elements should be arranged on the screen. The layout can determine where the \texttt{nodes} should be positioned or how the \texttt{edges} should be routed -- edge bundling. We attributed guidelines determining where a visualization part is displayed on the screen to the \texttt{layout} category, as determining where something should be positioned is the core task of a layout. Guidelines on layouts for node-link diagrams which we reviewed and structured are i.a. those: \cite{Pohl09,Didimo14,Kieffer2016,6065011,6596142,4658147,Telea09}

Since the node-link diagram's elements are \texttt{nodes} and \texttt{edges}, there are also guidelines for this. For the edges, there are \texttt{directed edges} and \texttt{undirected edges} according to the graph type. In addition, \texttt{additional information} can also be displayed on nodes and edges. There are i.a. guidelines for visualizing \texttt{group information} or \texttt{multivariate data} on edges.\\
Guidelines which we incorporate in our structure are for instance those:
\begin{itemize}
	\item Nodes i.a. \cite{tennekes2014tree}
	\item Edges
	\begin{itemize}
		\item Directed edges i.a. \cite{10.1145/1518701.1519054,5742390}
		\item Undirected edges i.a. \cite{10.1007/978-3-642-31223-6_34,10.1007/978-3-642-36763-2_40}
	\end{itemize}
	\item Additional information
	    \begin{itemize}
        	\item Group information i.a. \cite{Jianu14,6876036}
	        \item Multivariate data i.a. \cite{Schoeffel16}
	     \end{itemize}
\end{itemize}

Since humans look at node-link diagrams, their readability is also pivotal. Consequently, there is a huge pile of guidelines on graph readability resp. graph aesthetics. For example, edge crossings should be avoided \cite{Huang07}. Further guidelines on graph readability resp. aesthetics reviewed and strctured are i.a. these: \cite{Purchase:1997a,Ware02,Huang07_2,Huang09}

The graphs can also be \texttt{dynamic}. Dynamic graphs are graphs which change over time. Usually, they are recorded over multiple consecutive time steps. Guidelines for dynamic graphs i.a. specify whether the graphs should be displayed with a fixed layout to ensure the \texttt{mental map}'s stability \cite{misue1995layout}. The mental map is the mental image of the dynamic graph and its structure which humans use for thinking and reasoning -- e.g., to identify what changed in the graph and, consequently, how similar the graphs of both time steps are. For dynamic graphs, there are also guidelines which determine how to visualize their changes -- with animation or small multiples . While animation depicts all time steps one after the other like a video, small multiple visualize all time steps as small visualization in a matrix-like grid. Further guidelines on this topic which we reviewed are: \cite{purchase2007important,10.1007/978-3-642-36763-2_42,Archambault12,Archambault:2011,Brandes11,Boyandin12,Archambault11,Farrugia11}.
\FloatBarrier

\subsection{Graph Type Perspective} 
\label{sub:graph_type_perspective}

\begin{figure}[tb]
  \centering
    \includegraphics[width=.6\textwidth]{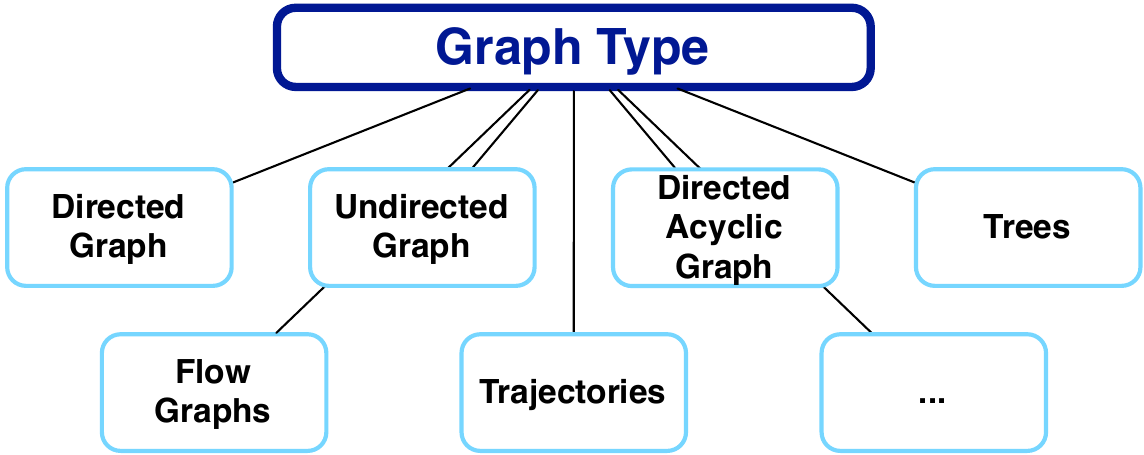}
  \caption{Graph type -- taxonomy perspective}
  \label{fig:pictures_DACH_Chapter3_Perspective_GraphType}
\end{figure}

Another way to systematize the guidelines is to structure them along to the type of graph they concern. Figure~\ref{fig:pictures_DACH_Chapter3_Perspective_GraphType} visualizes the structure of this perspective. This is a reasonable alternative perspective since generally only the guidelines developed for a particular graph type are relevant for that graph type. A prime example is the example of von Landesberger et al.'s \cite{Landesberger16} usage of tapered edges for geo-located networks which we presented in our introduction. The literal guideline recommendation is: if you want to visualize a directed graph as a node-link diagram, then use tapered edges to visualize the edges' direction. There is no more detail or differentiation which further criteria have to be met. When someone now tries, like von Landesberger et al. \cite{Landesberger16} did, to use the guideline for a more specific directed graph, in their case a geo-located directed graph, we can see that the guideline suddenly is not that suitable anymore. But why? It is due to the strong influence of the graph's type and the type-specific properties on the suitability of a visualization decision. For a network with no a priori defined node positions where a force-directed layout can push the nodes apart the tapered edge guideline is well suited, whereas for a geo-located graph with a priori defined node positions it leads to a huge amount of visual clutter.

Our system -- the GuidelineExplorer -- currently encompasses guidelines for the following graph types:
\begin{itemize}
	\item Undirected graphs
	\item Directed graphs
	\item Directed acyclic graphs
	\item Trees
	\item Flow graphs
	\item Trajectories
\end{itemize}
As the graph type perspective directly results from the graphs which are used in guideline formulating research, this perspective is directly extensible. The only check which is necessary for a new guideline which is to be classified based on our taxonomy is whether or not the graph type on which it is formulated is already present in the graph type perspective. In case the graph type is already present, the guideline can be directly classified and assigned to the other already present guidelines. In case the graph type is not present, the graph type has to be added to the graph type perspective and then the guideline can be classified and assigned.	
\FloatBarrier


\subsection{If-Type Perspective} 
\label{sub:if_type_perspective}

\begin{figure}[tb]
  \centering
    \includegraphics[width=.6\textwidth]{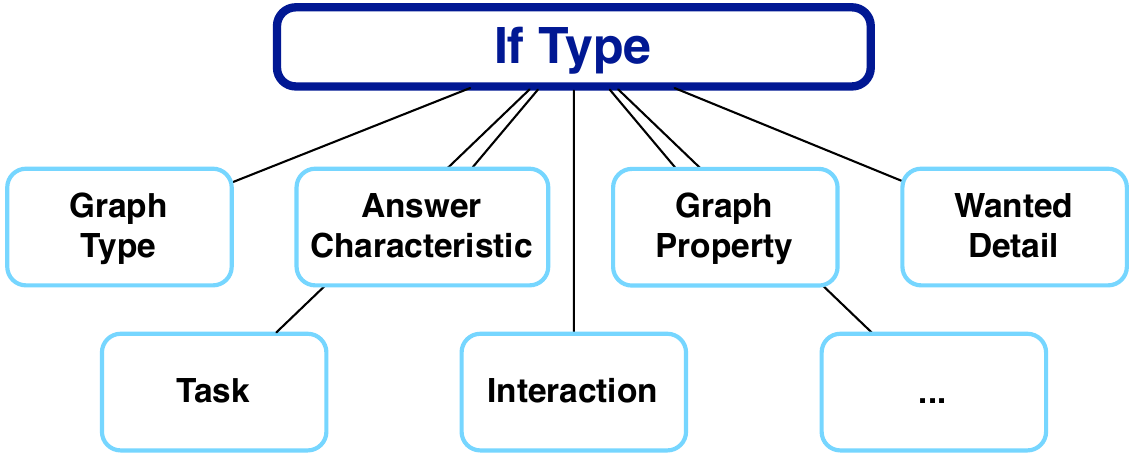}
  \caption{If-type -- taxonomy perspective}
  \label{fig:pictures_DACH_Chapter3_Perspective_IfType}
\end{figure}

A guideline is always structured in such a way that, if a certain condition applies, it tells you how the visualization should be designed. These conditions are of different types (if-condition types). Some if-conditions are based on the graph type as, for instance, these guidelines \cite{Didimo14,6065011,6596142,10.1145/1518701.1519054}. Still others depend on graph properties \cite{Ghoniem2005,Ghoniem04,doi:10.1057/palgrave.ivs.9500116} or characteristics of the answers of the users of the visualization \cite{5742390,10.1007/978-3-642-25878-7_22,Archambault11}. For example, the recommendations of the guidelines differ depending on whether the response of the visualization users should be correct or fast or as correct as possible.

A visualization designer usually knows the if-condition types because she knows the properties and type of her graph and also what is important when using her visualization; e.g., fast answers. This means that structuring by the if-condition types links what the visualization designer knows with what she currently does not know -- the most suitable visualization decision.
In the rare case that she does not know the afore mentioned if-conditions, this taxonomy perspective can help her to 
\begin{enumerate}
	\item pay attention to these important aspects of visualization design manifested in the if-conditions of guidelines,
	\item characterize what the concrete manifestation of these aspects are for her concrete case,
	\item consequently, make the right visualization decision based on the guidelines suitable for her data and visualization goal.
\end{enumerate} 

The if-types which our if-type perspective encompass are these (cf. Figure~\ref{fig:pictures_DACH_Chapter3_Perspective_IfType}):
\begin{itemize}
	\item Graph type -- e.g., trees, directed acyclic graphs, or (un)directed graphs
	\item Answer characteristic -- e.g., speed or correctness of the visualization users' answers
	\item Graph property -- e.g., the graph is small, large, or dense
	\item Wanted detail -- e.g., whether colors or curved edges are wished for
	\item Task -- i.e., the task which the visualization should support; e.g., finding the shortest path
	\item Interaction -- i.e.,  whether a specific interaction with a visualization is possible; e.g., zooming
\end{itemize}
Certainly, the if-types depend on the respective guidelines as different guidelines have different types of if-conditions -- e.g., answer characteristics, graph type, or task. But, as we identified the if-conditions' types with a thematic analysis \cite{doi:10.1191/1478088706qp063oa,thematicAnalysis}, the categories of the if-types are easily extensible.
\FloatBarrier

\subsection{Task Perspective} 
\label{sub:task_perspective}

\begin{figure}[tb]
  \centering
    \includegraphics[width=\textwidth]{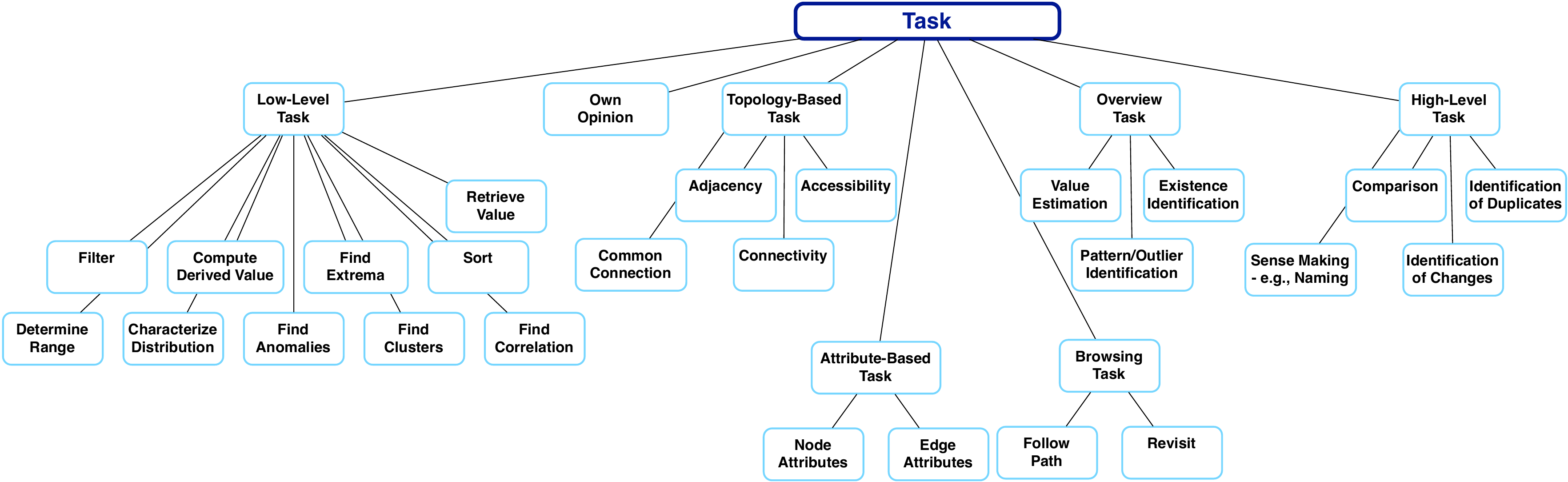}
  \caption{Task -- taxonomy perspective}
  \label{fig:pictures_DACH_Chapter3_Perspective_Task}
\end{figure}

The tasks to be solved have a considerable influence on the suitability of visualization decisions. Among others, this is made very clear by the work of Munzner \cite{5290695} and Miksch and Aigner \cite{MIKSCH2014286}. Each guideline is also examined on the basis of specific tasks and finally formulated. This means that the final guideline and the tasks on the basis of which the guideline was formulated cannot be separated. Thus, the tasks used in the investigation and the guidelines' final formulation are also a useful taxonomical perspective. This perspective can help to check whether the guideline suitable for the visualization currently being created has been investigated and formulated based on the task to be supported by the visualization. Knowing this helps the visualization designer to assess whether difficulties with the visualization decision from the guideline are likely to occur due to the tasks to be supported or whether this is unlikely, since the guideline was formulated based on the to be supported tasks.

As a basis for our task perspective, we chose the graph visualization task taxonomy of Lee et al. \cite{10.1145/1168149.1168168}. We decided to use the work of Lee et al. as it is a general taxonomy as compared to those on dynamic graphs \cite{doi:10.1111/cgf.12791,eurovisstar.20141174}, evolution analysis \cite{6620874}, group-level graph visualization \cite{saket2014group,vehlow15state}, or multi-faceted graphs \cite{eurovisstar.20151109}.\\
The work of Lee et al. \cite{10.1145/1168149.1168168} encompasses the following tasks:
\begin{enumerate}
	\item \textbf{Low-Level Tasks}
	\begin{enumerate}
		\item Retrieve value\\
		Given a set of nodes and/or edges, find the value of the additional information or multivariate data visualized for this set of nodes and/or edges
		\item Filter\\
		Given a set of nodes and/or edges and a set of conditions on the visualized additional information or multivariate data, find the nodes and/or edges satisfying those conditions
		\item ``Compute'' derived value\\
		Given a set of nodes and/or edges, ``compute'' an aggregating numeric value of those nodes and/or edges; e.g., mean/median or count
		\item Find extrema\\
		Given a set of nodes and/or edges and additional information or multivariate data visualized for this set of nodes and/or edges, find nodes and/or edges possessing extreme values of the visualized additional information or multivariate data
		\item Sort\\
		Given a set of nodes and/or edges and additional information or multivariate data visualized for this set of nodes and/or edges, sort them according to an ordinal metric
		\item Determine range\\
		Given a set of nodes and/or edges and additional information or multivariate data visualized for this set of nodes and/or edges, find the value range of an attribute of interest of the additional information resp. the multivariate data visualized for the given set of nodes and/or edges
		\item Characterize distribution\\
		Given a set of nodes and/or edges and a quantitative attribute of interest of the visualized additional information resp. multivariate data for the given set of nodes and/or edges, characterize the attribute of interest's distribution over the given set of nodes and/or edges
		\item Find anomalies\\
		Find anomalies in a given set of nodes and/or edges and visualized additional information resp. multivariate data for this set of nodes and/or edges with respect to a specific relationship or expectation; e.g., statistical outliers
		\item Find clusters\\
		Given a set of nodes and/or edges, find clusters of similar nodes and/or edges.
		\item Find correlations\\
		Given a set of nodes and/or edges and additional attributes of this set of nodes and/or edges, determine useful correlations between the values of those attributes.
	\end{enumerate}
	\item \textbf{Topology-Based Tasks}
	\begin{enumerate}
		\item Adjacency -- direct connection\\
		Determine i.a. whether a node is a neighbor of a certain node or the number of neighbors of a certain node.
		\item Accessibility -- (in)direct connection\\
		Determine i.a. which nodes are accessible via a certain node or the number of accessible nodes -- also based on distance constraints; e.g., distance less than n.
		\item Common connection\\
		Determine a set of nodes which are connected to each other.
		\item Connectivity\\
		Determine the shortest path between nodes, clusters, connected components, bridges and more.
	\end{enumerate}
	\item \textbf{Attribute-Based Tasks}
	\begin{enumerate}
		\item Node attributes\\
		Find a set of nodes having certain attributes or review a certain set of nodes with respect to their attributes.
		\item Edge attributes\\
		Find a set of edges having certain attributes or review a certain set of edges with respect to their attributes.
	\end{enumerate}
	\item \textbf{Browsing Tasks}
	\begin{enumerate}
		\item Follow path\\
		A predefined path needs to be followed or searched for.
		\item Revisit\\
		A previously visited node needs to be revisited -- e.g., find another animal which uses the same food source as the first.
	\end{enumerate}
	\item \textbf{Overview Tasks}\\
	Overview tasks encompass i.a.:
	\begin{enumerate}
		\item Estimate the value of a specific property -- e.g., the graph's size
		\item Existence of clusters or connected components
		\item Pattern or outlier identification
	\end{enumerate}
	\item \textbf{High-Level Tasks}\\
	High-level tasks encompass i.a.
	\begin{enumerate}
		\item Comparison
		\item Identification of duplicates
		\item Sense making tasks -- e.g., give a cluster a meaningful name
		\item Identification of changes
	\end{enumerate}
\end{enumerate}
We added the task category ``own opinion'' as, in the field of guideline research, researchers often ask their participants about their own opinion --i.e., which design they liked most or helped them best resp. was most suited for a specific task. Figure~\ref{fig:pictures_DACH_Chapter3_Perspective_Task} shows the structure of our task perspective.

The task perspective is extensible as well. For instance, the high-level or overview tasks which are covered in a rather general sense by Lee et al. \cite{10.1145/1168149.1168168} are covered by more specialized taxonomies -- e.g., for temporal or dynamic graphs (cf. e.g., \cite{7091028,doi:10.1111/cgf.12791,eurovisstar.20141174})-- in more detail. So, these taxonomies could be used as extensions of Lee et al.'s. In case there will be guideline formulating studies in the future which start using high-level tasks, it is advisable to implement such extensions of the task perspective of our taxonomy.
\FloatBarrier

\section{System Presentation} 
\label{sec:guidelineexplorer_system_description}

\begin{figure}[tb]
  \centering
    \includegraphics[width=\textwidth]{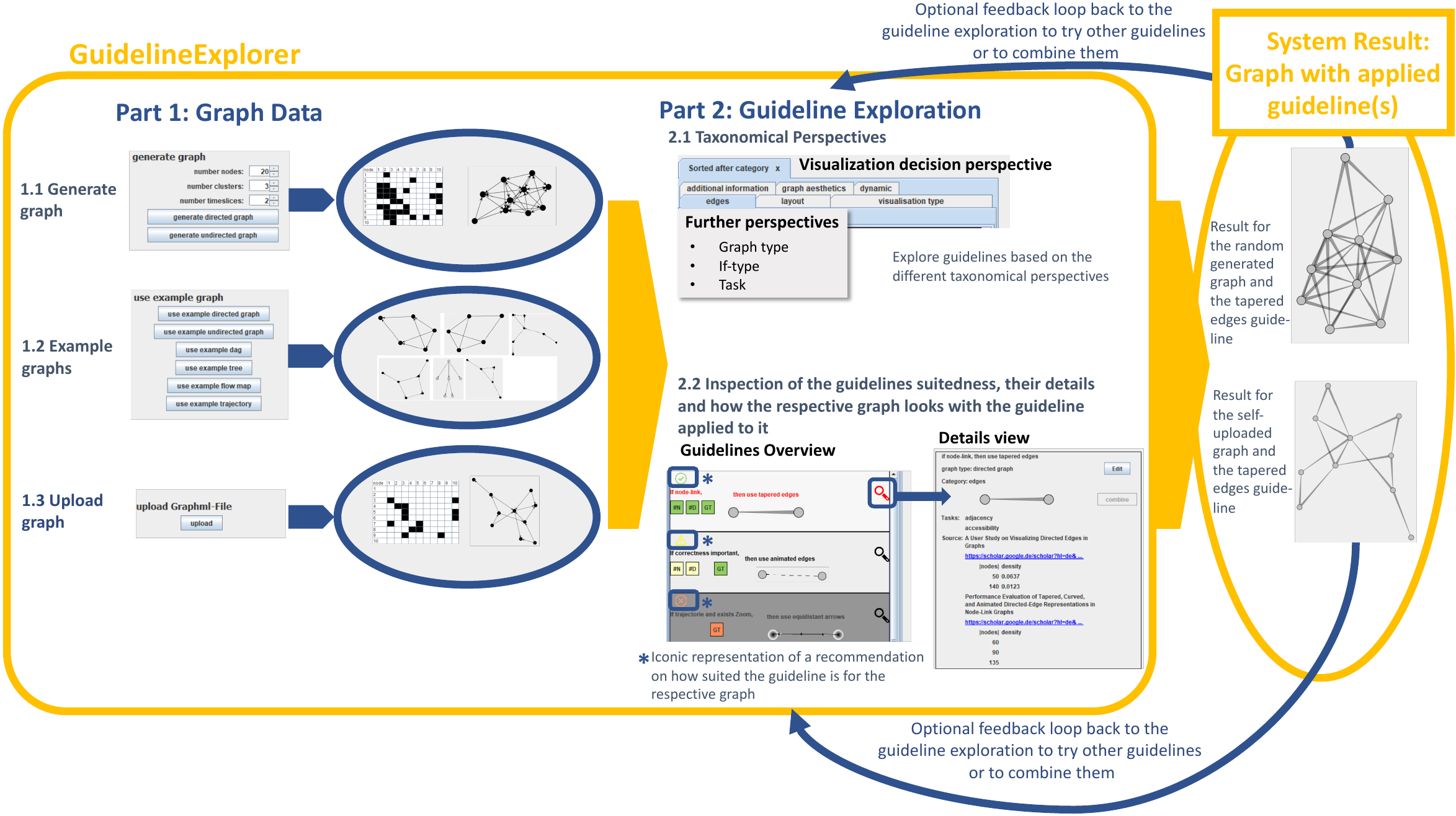}
  \caption[GuidelineExplorer -- system schema]{Schema of the GuidelineExplorer system. First, the visualization designer can generate a graph (1.1), use example graphs (1.2), or upload her own graph (1.3) to explore possibly suitable guidelines for the respective graph (Part 1). Then, cf. Part 2, the visualization designer can explore the guidelines based on the different taxonomical perspectives (2.1). GuidelineExplorer presents the respective guidelines in a list-based guidelines overview where details to a certain guideline can be acquired on demand. In case the visualization designer is not satisfied with the appearance of her graph with the respective guideline applied, she can loop back and explore further guidelines (System Result, Optional Feedback Loop).}
  \label{fig:pictures_DACH_Chapter3_GuidelineExplorer_Schema}
\end{figure}

Our system -- GuidelineExplorer -- consists of two parts (cf. Figure~\ref{fig:pictures_DACH_Chapter3_GuidelineExplorer_Schema}):
\begin{enumerate}
	\item \textbf{Graph data view}\\
	Here, the visualization designer can generate a graph (cf. Figure~\ref{fig:pictures_DACH_Chapter3_GuidelineExplorer_Schema} -- \texttt{1.1}), use one of the example graphs (cf. Figure~\ref{fig:pictures_DACH_Chapter3_GuidelineExplorer_Schema} -- \texttt{1.2}), or upload a graph of her own, cf. Figure~\ref{fig:pictures_DACH_Chapter3_GuidelineExplorer_Schema} -- \texttt{1.3}, to explore suitable guidelines for the chosen graph.
	\item \textbf{Guideline exploration view}\\
	This view presents the guidelines based on the taxonomical perspectives which we presented in Section~\ref{sec:taxonomicalPerspectives} (cf. Figure~\ref{fig:pictures_DACH_Chapter3_GuidelineExplorer_Schema} -- \texttt{2.1}). GuidelineExplorer shows only those categories of the taxonomical perspectives for which it has implemented guidelines. For instance, in case there are currently no guidelines implemented for undirected edges, GuidelineExplorer will only show the structure \texttt{Edges} $\rightarrow$ \texttt{directed} in spite of the actual taxonomy has the structure \texttt{Edges} $\rightarrow$ \texttt{directed}, \texttt{undirected}. We decided for this to clearly separate the actual taxonomical perspective and which guidelines are currently implemented in GuidelineExplorer.  Initially, the visualization designer sees the foundational perspective of visualization decisions and the guidelines in a scrollable overview (cf. Figure~\ref{fig:pictures_DACH_Chapter3_GuidelineExplorer_Schema} -- \texttt{2.2}, \texttt{Guidelines Overview}). The overview shows a short version of the guideline in an ``if-then''-statement form. It also gives iconic recommendations on the suitability of the respective guideline for the chosen graph based on the graph type (\texttt{GT}), the graph's number of nodes (\texttt{\#N}) and density (\texttt{\#D}); well suited -- graph type, number of nodes and density match (\includegraphics[width=0.35cm]{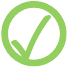}), medium -- graph type matches but number of nodes and density not (\includegraphics[width=0.35cm]{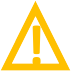}), not suited -- the graph type does not match (\includegraphics[width=0.35cm]{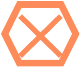}). In case the guideline is not suitable for the respective graph, the list cell is grayed out and the guideline can not be applied to the graph. Furthermore, the guideline exploration view provides a small visualization of how the guideline looks. It also offers the visualization designer to inspect details of the respective guideline -- i.a. the graph type it was formulated for, the tasks which were used to investigate the guideline, or the source paper via Google Scholar (cf. Figure~\ref{fig:pictures_DACH_Chapter3_GuidelineExplorer_Schema} -- \texttt{2.2}, \texttt{Details view}).
\end{enumerate}
Finally, via the \texttt{Guideline Exploration View} (cf. Section~\ref{sub:guideline_exploration_view}) the visualization designer can apply the guideline to the graph she generated, selected, or uploaded in the \texttt{Graph Data View} (cf. Section~\ref{sub:graph_data_view}). GuidelineExplorer directly visualizes the graph with the applied guideline. In case the visualization designer does not like the result, she can do a feedback loop back to the \texttt{Guideline Exploration View} and look for other guidelines. She can even experiment with combining guidelines. What does this mean? This means, the visualization designer can for instance try out the overloaded orthogonal layout \cite{Didimo14} which is originally visualized with arrow-based edges with tapered edges. Admittedly, this combining functionality goes considerably beyond the results of the respective guideline papers we collected resp. implemented as there are no empirical results on how to combine different guidelines yet. However, we were convinced that such a functionality supports design creativity and maybe even reveals interesting designs to be studied in an empirical study.

Our contributions to this work is not only the implementation of the guidelines -- on the contrary -- our main contributions are the proposed taxonomical perspectives, the transfer of the taxonomy into our system, the implementation of the system as well as the implementation of an initial subset of guidelines to be able to show the benefits of our system at work.
\FloatBarrier

\subsection{Graph Data View} 
\label{sub:graph_data_view}

\begin{figure}[tb]
  \centering
    \includegraphics[width=.7\textwidth]{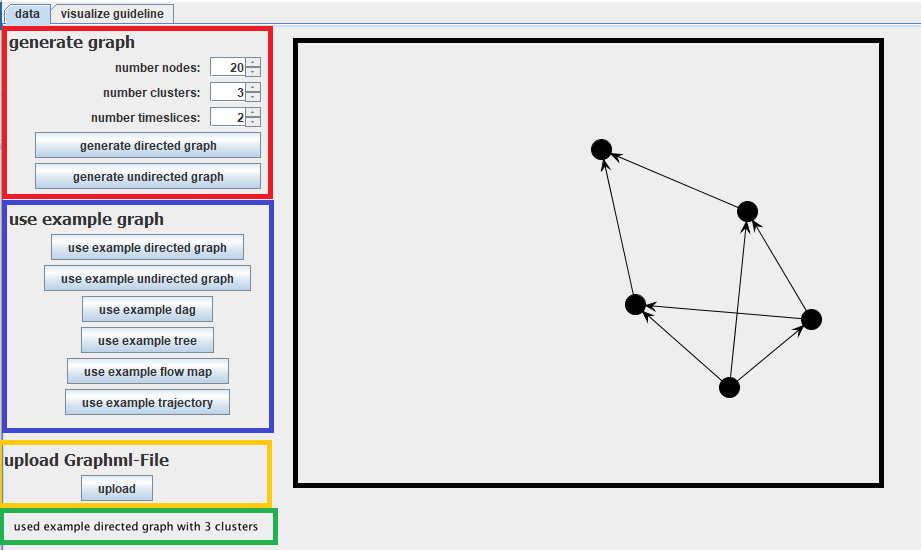}
  \caption[GuidelineExplorer -- graph data view]{Graph data view -- here, the visualization designer can decide whether she wants to use a generated graph, one of the example graphs, or one of her own graphs to explore possibly suitable guidelines with GuidelineExplorer (Figure taken from \cite{LisaBA}).}
  \label{fig:pictures_DACH_Chapter3_GuidelineExplorer_graphDataView}
\end{figure}

\begin{figure}[tb]
  \centering
    \includegraphics[width=.9\textwidth]{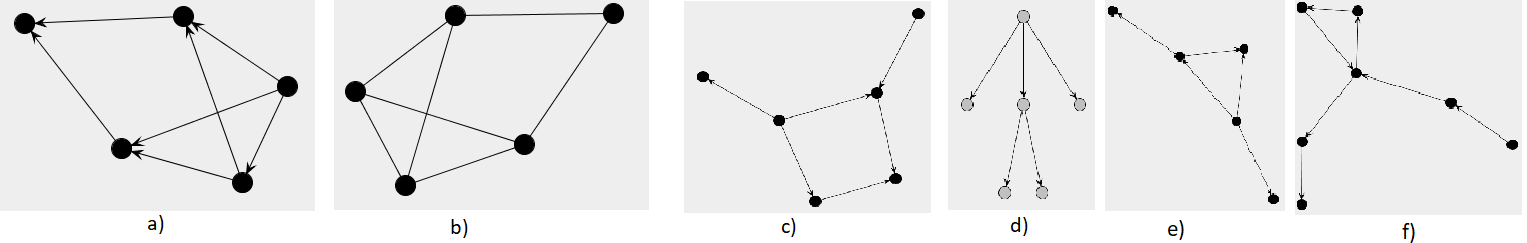}
  \caption[GuidelineExplorer -- example graphs]{Example graphs of GuidelineExplorer -- a) directed graph, b) undirected graph, c) directed acyclic graph (DAG), d) tree, e) flow map, f) trajectory (Figure taken from \cite{LisaBA})}
  \label{fig:pictures_DACH_Chapter3_GuidelineExplorer_exampleGraphs}
\end{figure}

In the graph data view, the visualization designer selects the graph for which she wants to explore suitable guidelines for. She can:
\begin{enumerate}
	\item generate a graph (cf. Figure~\ref{fig:pictures_DACH_Chapter3_GuidelineExplorer_graphDataView} -- $\color{lisaRed}{\Box}$),
	\item use one of the example graphs (cf. Figure~\ref{fig:pictures_DACH_Chapter3_GuidelineExplorer_graphDataView} -- $\color{lisaBlue}{\Box}$,~\ref{fig:pictures_DACH_Chapter3_GuidelineExplorer_exampleGraphs}), or
	\item upload a graph of her own as a GraphML\footnote{http://graphml.graphdrawing.org} file (cf. Figure~\ref{fig:pictures_DACH_Chapter3_GuidelineExplorer_graphDataView} -- $\color{lisaYellow}{\Box}$).
\end{enumerate}
The respective graph is described with a short textual description explaining its number of clusters, that it is an example graph or its number of nodes and edges (cf. Figure~\ref{fig:pictures_DACH_Chapter3_GuidelineExplorer_graphDataView} -- $\color{lisaGreen}{\Box}$) and visualized (cf. Figure~\ref{fig:pictures_DACH_Chapter3_GuidelineExplorer_graphDataView} -- $\color{black}{\Box}$).
%
\FloatBarrier

\paragraph{Generate Graph.} 
\label{par:generate_graph}
For the graph generation GuidelineExplorer expects the number of nodes as an input. The visualization designer has further the option to determine further graph characteristics  -- the number of clusters and time slices. Based on this input, the visualization designer has the option to generate either an undirected or a directed graph with the Barabasi-Albert\footnote{http://jung.sourceforge.net/doc/api/edu/uci/ics/jung/algorithms/generators/random/BarabasiAlbertGenerator.html} generator implementation of the JUNG framework\footnote{http://jung.sourceforge.net}.
 
\paragraph{Example Graphs.} 
\label{par:example_graphs}
To allow the visualization designer to use all graph types for which GuidelineExplorer encompasses implemented guidelines, we added example graphs (cf. Figure~\ref{fig:pictures_DACH_Chapter3_GuidelineExplorer_exampleGraphs}).
\FloatBarrier

\paragraph{Upload Graph.} 
\label{par:upload_graph}
It is also possible to upload a graph in the form of a GraphML file. After clicking on the upload button (cf. Figure \ref{fig:pictures_DACH_Chapter3_GuidelineExplorer_graphDataView} -- $\color{lisaYellow}{\Box}$) a window opens where the GraphML file can be selected. Its ``edgedefault'' field determines whether the uploaded graph is undirected or directed.

\subsection{Guideline Exploration View} 
\label{sub:guideline_exploration_view}	

\begin{figure}[tb]
  \centering
    \includegraphics[width=\textwidth]{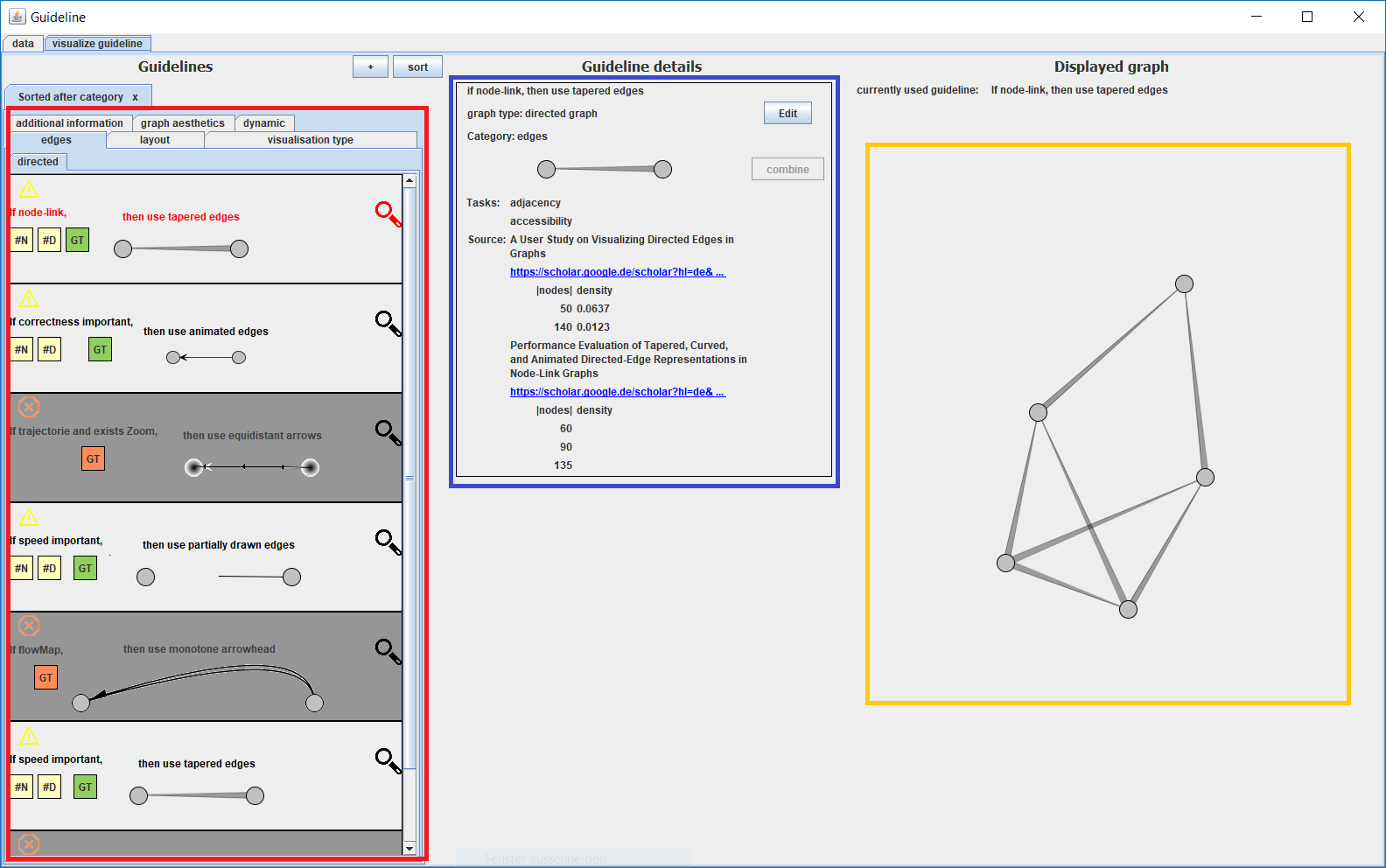}
  \caption[GuidelineExplorer -- guideline exploration view]{In GuidelineExplorer's guideline exploration view, the visualization designer can explore the guidelines with respect to their suitability to her respective graph (Figure taken from \cite{LisaBA}). The guideline exploration view is divided into the guidelines overview ($\color{lisaRed}{\Box}$ ,cf. Section~\ref{ssub:guidelines_overview}), the details view ($\color{lisaBlue}{\Box}$) and the graph visualized with guidelines applied ($\color{lisaYellow}{\Box}$) (cf. Section~\ref{ssub:details_view_and_graph_visualized_with_applied_guideline}).}
  \label{fig:pictures_DACH_Chapter3_GuidelineExplorer_guidelineExplorationView}
\end{figure}

\begin{figure}[tb]
  \centering
    \includegraphics[width=.6\textwidth]{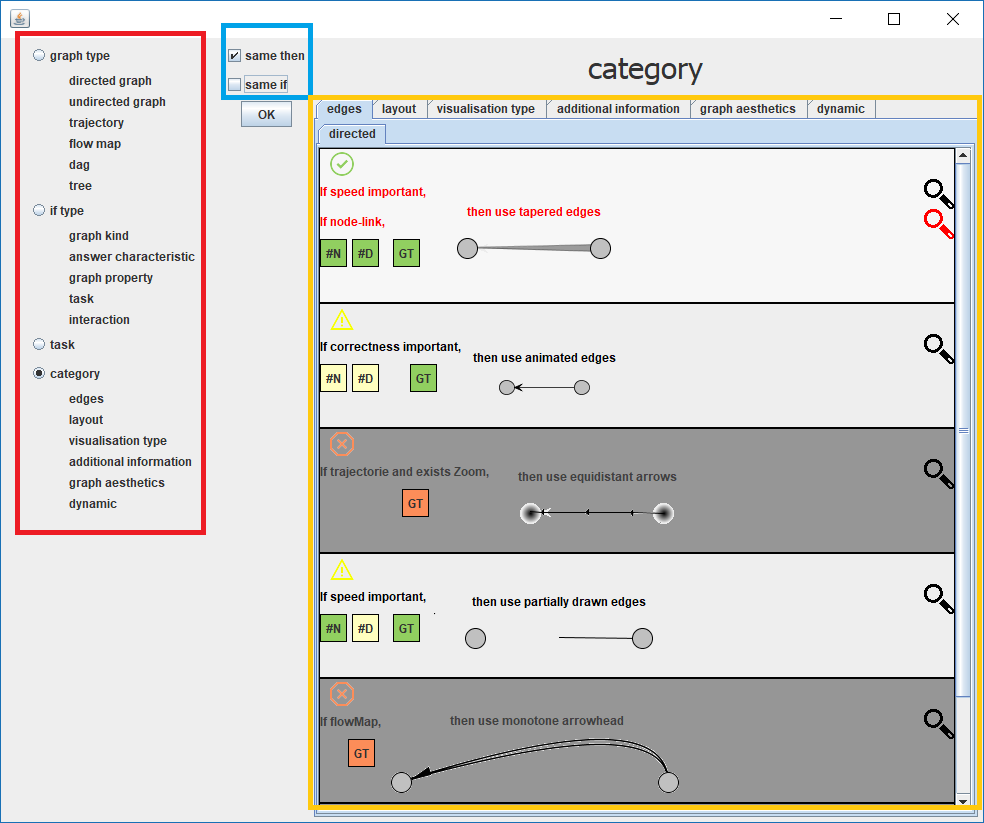}
  \caption[GuidelineExplorer -- GUI for changing the taxonomical perspectives]{Sort-Button (Figure taken from \cite{LisaBA}) -- Here, the visualization designer can change the taxonomical perspective via choosing one of the other perspectives from the radio button menu ($\color{lisaRed}{\Box}$, cf. Sections~\ref{sub:basic_taxonomy}-\ref{sub:task_perspective}). Further, she can decide whether she likes to group the guidelines with respect to the same ``if''- or the same ``then''-statement ($\color{lisaBlue2}{\Box}$). This supports the visualization designer i.a. in identifying a case when researchers came to different conclusions -- i.e., here, different ``then''-statements. The visualization designer can also see, directly in the pop-up, a preview of the new structure of the guidelines overview ($\color{lisaYellow}{\Box}$) and she can inspect details of the guidelines from there (magnifying glass).}
  \label{fig:pictures_DACH_Chapter3_sortButtonPopUp}.
\end{figure}

Here, the visualization designer can explore the guidelines based on the taxonomical perspectives introduced in Section~\ref{sec:taxonomicalPerspectives}. The initial perspective is the one on the visualization decisions (foundational perspective, cf. Section~\ref{sub:basic_taxonomy}). The visualization designer can switch perspectives via the \texttt{sort}-button. Via this triggered pop-up, see Figure~\ref{fig:pictures_DACH_Chapter3_sortButtonPopUp}, she can select one of the other perspectives ($\color{lisaRed}{\Box}$) -- \texttt{graph type} (cf. Section~\ref{sub:graph_type_perspective}), \texttt{if-type} (cf. Section~\ref{sub:if_type_perspective}), \texttt{task} (cf. Section~\ref{sub:task_perspective}), \texttt{category} -- i.e., visualization decision (cf. Section~\ref{sub:basic_taxonomy}). The visualization designer can also see, directly in the pop-up, a preview of the new structure of the guidelines overview ($\color{lisaYellow}{\Box}$), she can inspect details of the guidelines from there (\texttt{magnifying glass}), and configure whether the guidelines should be grouped according to the ``if''- (\texttt{same if}) or the ``then'' (\texttt{same then}) part of the ``if-then''-statement ($\color{lisaBlue2}{\Box}$). If the visualization designer confirms a selection of one of the other perspectives, this taxonomical perspective is provided in addition to the foundational perspective. We find that providing different taxonomical perspectives side by side supports the visualization designer in getting to know the currently existing guideline research with its different facets. The option for grouping the guidelines according to the same ``if'' or the same ``then'' statement allows the visualization designer to investigate where research comes to a different result (=grouping according to the same ``if'') or where different conditions -- ```if''- part of the ``if-then'' statement -- lead to the same result (= grouping according to the same ``then''). We divided the guideline exploration view in three parts (cf. Figure~\ref{fig:pictures_DACH_Chapter3_GuidelineExplorer_guidelineExplorationView}):
\begin{enumerate}
	\item the guideline overview ($\color{lisaRed}{\Box}$),
	\item the details view ($\color{lisaBlue}{\Box}$), and
	\item the visualization of graph which the visualization designer selected in the graph data view (cf. Section~\ref{sub:graph_data_view}) with the guideline applied to it ($\color{lisaYellow}{\Box}$).
\end{enumerate}
We introduced the guidelines overview, to allow the visualization designer to quickly get an overview over the available guidelines and their suitability for her respective graph. In case a guideline attracts her interest, the details view allows for further inspections. We separated the guideline exploration view and the details view to not overload the former. As the final visualized graph with the guideline applied is again another aspect, we gave it also a separate view. With this multiple view approach we follow the recommendation of Plumlee et al. \cite{Plumlee:2006:ZVM:1165734.1165736}.

The users of GuidelineExplorer can even add new guidelines with the \texttt{+}-button. It opens a popup where all information on the guideline can be entered. When the user confirms her input a guideline class is generated which then has to be be implemented by the user so that the guideline gets its real functionality. Within GuidelineExplorer, the guideline is visible directly after the suer confirms her input.
\FloatBarrier

\subsubsection{Guidelines Overview} 
\label{ssub:guidelines_overview}	

\begin{figure}[tb]
  \centering
    \includegraphics[width=.75\textwidth]{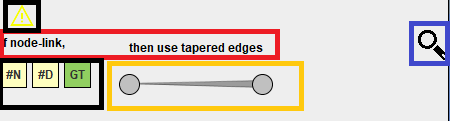}
  \caption[GuidelineExplorer -- single guideline]{In GuidelineExplorer a single guideline consists of a recommendation on the suitability of the guideline for the graph currently in use ($\color{black}{\Box}$) based on the graph's type (GT), the graph size in number of nodes (\#N), and the graph's density (\#D) ($\color{black}{\Box}$). Further, it consists of a short ``if-then''-statement expressing the recommendation of the guideline ($\color{lisaRed}{\Box}$) and a small visualization of how the guideline would look like applied to a graph ($\color{lisaYellow}{\Box}$). Finally, there is the option to inspect further details of the guideline in the details view ($\color{lisaBlue}{\Box}$)  (Figure taken from \cite{LisaBA}).}
  \label{fig:pictures_DACH_Chapter3_GuidelineExplorer_singleGuideline}
\end{figure}

The guidelines overview shows a list of guidelines available in GuidelineExplorer. In GuidelineExplorer, each guideline consists of the following parts (cf. Figure~\ref{fig:pictures_DACH_Chapter3_GuidelineExplorer_singleGuideline}):
\begin{enumerate}
	\item Recommendation on the suitability ($\color{black}{\Box}$)
	\begin{enumerate}
		\item A summarization with an iconic representation
		\begin{itemize}
			\item Well suited -- graph type, number of nodes and density match (\includegraphics[width=0.35cm]{pictures/DACH/Chapter3/icon_optimal.pdf})
			\item Medium -- graph type matches but number of nodes and density not (\includegraphics[width=0.35cm]{pictures/DACH/Chapter3/icon_medium.pdf})
			\item Not suited --the graph type does not match (\includegraphics[width=0.35cm]{pictures/DACH/Chapter3/icon_bad.pdf})
		\end{itemize}
		\item Detail information which consider the 
		\begin{itemize}
			\item Graph type (GT)
			\item Number of nodes of the graph (\#N)
			\item Graph density (\#D)
		\end{itemize}
		The detail information considers these graph characteristics as these have a remarkable influence on the suitability of a guideline for a respective graph. The example in the introduction or the use case show lively examples of this. It is only possible to apply the guideline if the graph type matched. In the other case, the application of a guideline would make no sense. For instance, applying a guideline for a directed graph to an undirected graph is highly likely to provoke misconceptions. An undirected graph visualized, for instance, with tapered edges could provoke the impression that the graph is actually a directed graph. In case of a graph type missmatch the detail information only shows the graph type box with an orange background (\fcolorbox{black}{colorbrewerorange}{GT}) and the cell is grayed out. In case the graph type matches (\fcolorbox{black}{colorbrewergreen}{GT}), GuidelineExplorer checks the selected graph's number of nodes and density with those of the graphs the guideline was investigated with. In case they match, it is likely that the guideline is straight forward applicable to the graph of the visualization designer. Consequently, the graph type and the number of nodes and density symbol are visualized with a green background (\fcolorbox{black}{colorbrewergreen}{\#N}, \fcolorbox{black}{colorbrewergreen}{\#D}, \fcolorbox{black}{colorbrewergreen}{GT}). In case either of those or both do not match , the guideline is still applicable to the visualization designer's graph as the graph type match, but it can happen that the visualization decision recommended by the guideline does not perform that well. So either of the symbols or both are visualized with a yellow background (\fcolorbox{black}{colorbreweryellow}{\#N}, \fcolorbox{black}{colorbreweryellow}{\#D}). Our use case shows such an example. Here, the tapered edge guideline of Holten et al. \cite{10.1145/1518701.1519054} is applied to a denser graph as compared to the graphs used for the guideline's investigation. Consequently, there is a lot of visual clutter leading to partially drawn edges \cite{10.1007/978-3-642-25878-7_22} performing better as they produce less visual clutter.
	\end{enumerate}
	\item A short version of the guideline in an ``if-then''-statement form ($\color{lisaRed}{\Box}$)\\
	We chose a short version of the guidelines to concisely communicate the guideline's recommendations. The visualization designer can inspect details in the details view or the paper which she can access via the details view. In case the visualization designer applied a guideline to a graph of hers, the ``if-then''-statement is depicted in red (cf. Figure~\ref{fig:pictures_DACH_Chapter3_GuidelineExplorer_guidelineExplorationView} -- \texttt{first list cell}). 
	\item A small visualization of how the guideline looks when being applied to a graph ($\color{lisaYellow}{\Box}$)
	\item Magnifying glass -- it opens the details view for the respective guideline ($\color{lisaBlue}{\Box}$)\\
	The magnifying glass of a specific guideline is depicted in red, if the specific guideline's details are currently loaded in the details view.
\end{enumerate}
\FloatBarrier

\subsubsection{Details View and Graph Visualized with Guidelines Applied} 
\label{ssub:details_view_and_graph_visualized_with_applied_guideline}

\begin{figure}[tb]
  \centering
    \includegraphics[width=.6\textwidth]{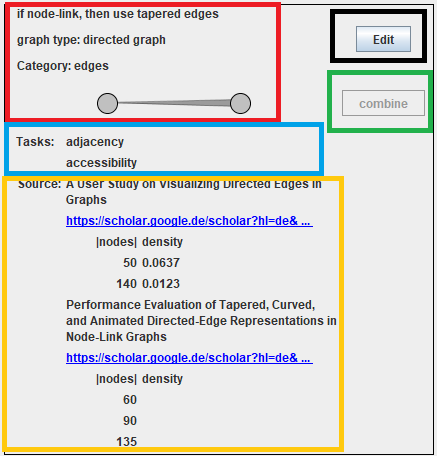}
  \caption[GudelineExplorer -- details view]{In the details view the visualization designer can inspect details of the respective guideline which she chose to inspect from the guidelines overview. The details view again shows the short ``if-then-statement'' expressing the recommendation of the guideline and also the visualization decision which is affected by the guideline ($\color{lisaRed}{\Box}$). Furthermore, it shows the task based on which the guideline was researched ($\color{lisaBlue2}{\Box}$). Finally, the details view shows the source paper(s) together with a Google Scholar link to access the paper(s) and information on the used graphs' size (number of nodes) and their density ($\color{lisaYellow}{\Box}$) (Figure taken from \cite{LisaBA}).}
  \label{fig:pictures_DACH_Chapter3_GuidelineExplorer_detailsView}
\end{figure}

\paragraph{Details View.} 
\label{par:details_view}
The details view is structured as follows (cf. Figure~\ref{fig:pictures_DACH_Chapter3_GuidelineExplorer_detailsView}): First there is again the short version of the guideline in an ``if-then''-statement form together with information on with which graph type the guideline was investigated, the visualization decision the guideline was formulated for, and a small visualization of how the guideline looks like when applied to a graph ($\color{lisaRed}{\Box}$). Furthermore, the details view shows which tasks were used to investigate the guideline. The list of tasks consists of all papers which contributed to the formulation of the guideline ($\color{lisaBlue2}{\Box}$). Finally, the details view shows the source paper(s) together with a Google Scholar link to access the paper(s) and information on the used graphs' size (number of nodes) and their density ($\color{lisaYellow}{\Box}$).

Via the details view, the visualization designer can either edit the respective selected guideline ($\color{black}{\Box}$) or use the functionality to combine guidelines ($\color{lisaGreen}{\Box}$).

\emph{\textbf{Combining Functionality.}} Herewith the visualization designer is able to combine guidelines. A main guideline must be selected, which is then combined with one or more other guidelines. To combine a guideline, the visualization designer must display the details of the guideline and use the \texttt{combine}-button. In general, guidelines can only be combined if:
\begin{enumerate}
	\item they are formulated for the same graph type and their graph type fits to the graph type of the graph currently selected by the visualization designer
	\item they do belong to a different taxonomical category as compared to the category of the main guideline or the other guidelines already combined with the main guideline
\end{enumerate}
Guidelines for different visualization types -- node-link diagrams and matrices -- are not combinable. As already explained, there are no empirical insights on how to combine guidelines. Nevertheless, it is clear that a combination of guidelines for two different visualization types makes no sense. For this functionality to work, all possible combinations must be anticipated and implemented in advance.

\begin{figure}[tb]
  \centering
    \includegraphics[width=.9\textwidth]{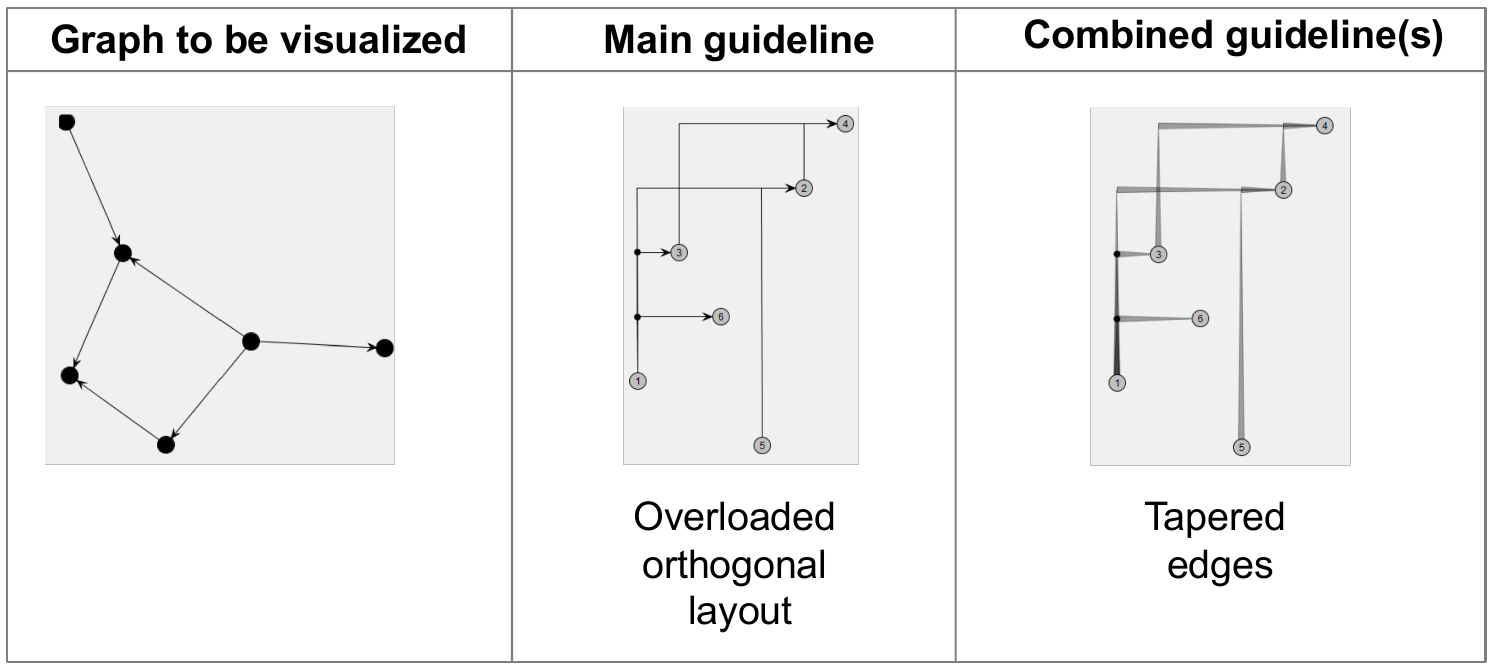}
  \caption[GuidelineExplorer -- combining functionality example 1]{Combining functionality example 1 -- combination of two guidelines: overloaded orthogonal layout (\cite{Didimo14}, main guideline) combined with tapered edges (\cite{10.1145/1518701.1519054,holten:hal-00696823}, combined guideline)}
  \label{fig:pictures_DACH_Chapter3_GuidelineCombinationExample1}
\end{figure}
	
\begin{figure}[tb]
  \centering
    \includegraphics[width=.9\textwidth]{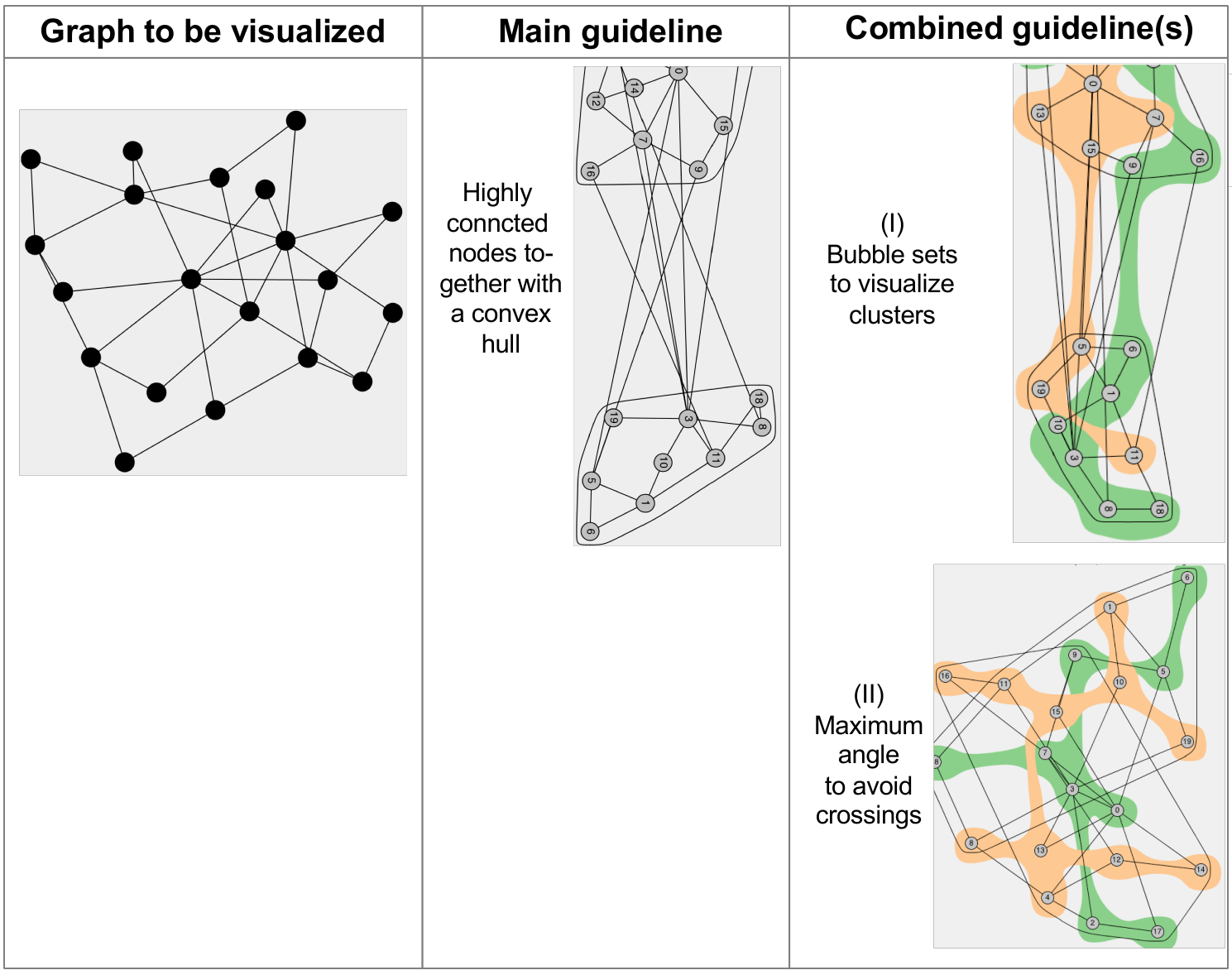}
  \caption[GuidelineExplorer -- combining functionality example 2]{Combining functionality example 2 -- combination of three guidelines: visualize highly connected nodes together with a convex hull (\cite{4658147}, main guideline) combined with (I) bubble sets to visualize clusters \cite{Jianu14} and (II) maximum angle of the edges and avoid edge crossings \cite{Ware02} ((I), (II) = combined guidelines).}
  \label{fig:pictures_DACH_Chapter3_GuidelineCombinationExample2}
\end{figure}
	
\begin{figure}[tb]
  \centering
    \includegraphics[width=.9\textwidth]{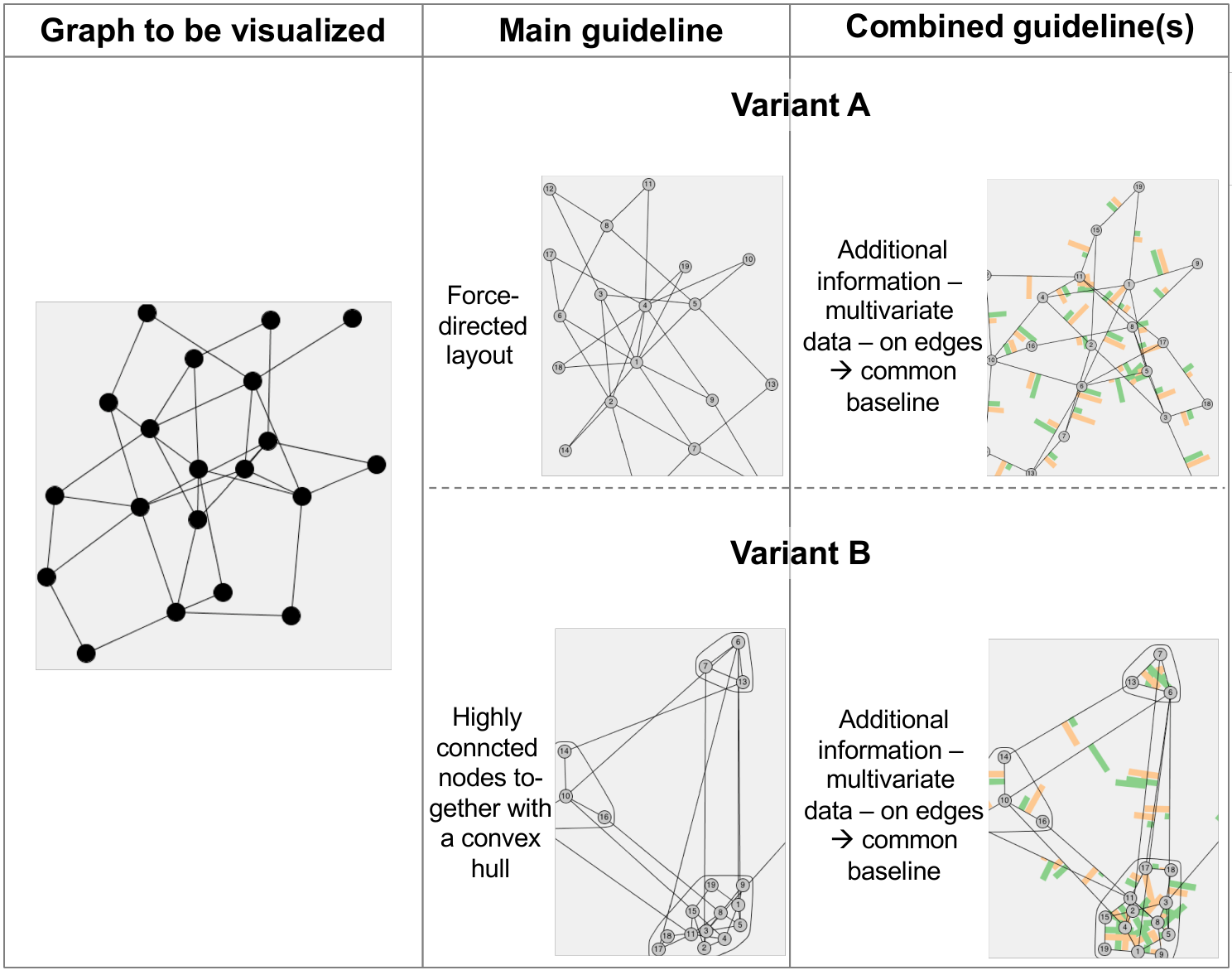}
  \caption[GuidelineExplorer -- combining functionality example 3]{Combining functionality example 3 -- combination of two guidelines in two variants (Variant A, Variant B): 1) for a node-link diagram use a force-directed layout \cite{Pohl09} or 2) visualize highly connected nodes together with a convex hull \cite{4658147} ( 1) or 2) = main guideline) combined with multivariate data visualized as on the edges as bar charts \cite{Schoeffel16} (combined guideline).}
  \label{fig:pictures_DACH_Chapter3_GuidelineCombinationExample3}
\end{figure}

Figure~\ref{fig:pictures_DACH_Chapter3_GuidelineCombinationExample1},~\ref{fig:pictures_DACH_Chapter3_GuidelineCombinationExample2}, and~\ref{fig:pictures_DACH_Chapter3_GuidelineCombinationExample3} show some examples of the visualization result based on the combination of different guidelines.
\FloatBarrier

\paragraph{Graph Visualized with Applied Guideline.} 
\label{par:graph_visualized_with_applied_guideline}
The visualized graph directly responds to the user interactions (cf. Figure~\ref{fig:pictures_DACH_Chapter3_GuidelineExplorer_guidelineExplorationView} -- $\color{lisaYellow}{\Box}$). It changes based on the currently selected guideline or the combination of guidelines (cf. Figure~\ref{fig:pictures_DACH_Chapter3_GuidelineExplorer_guidelineExplorationView} -- $\color{lisaYellow}{\Box}$). Furthermore, it displays the guideline currently applied in form of the short ``if-then''-statement.



\section{Use Case} 
\label{sec:use_case_guidelines}
Here, the visualization designer wants to visualize directed graph and she has the question of ``How to visualize the direction of the directed edges best?''. So, she keeps the foundational taxonomy perspective -- the visualization decisions.
\begin{figure}[tb]
  \centering
    \includegraphics[width=.4\textwidth]{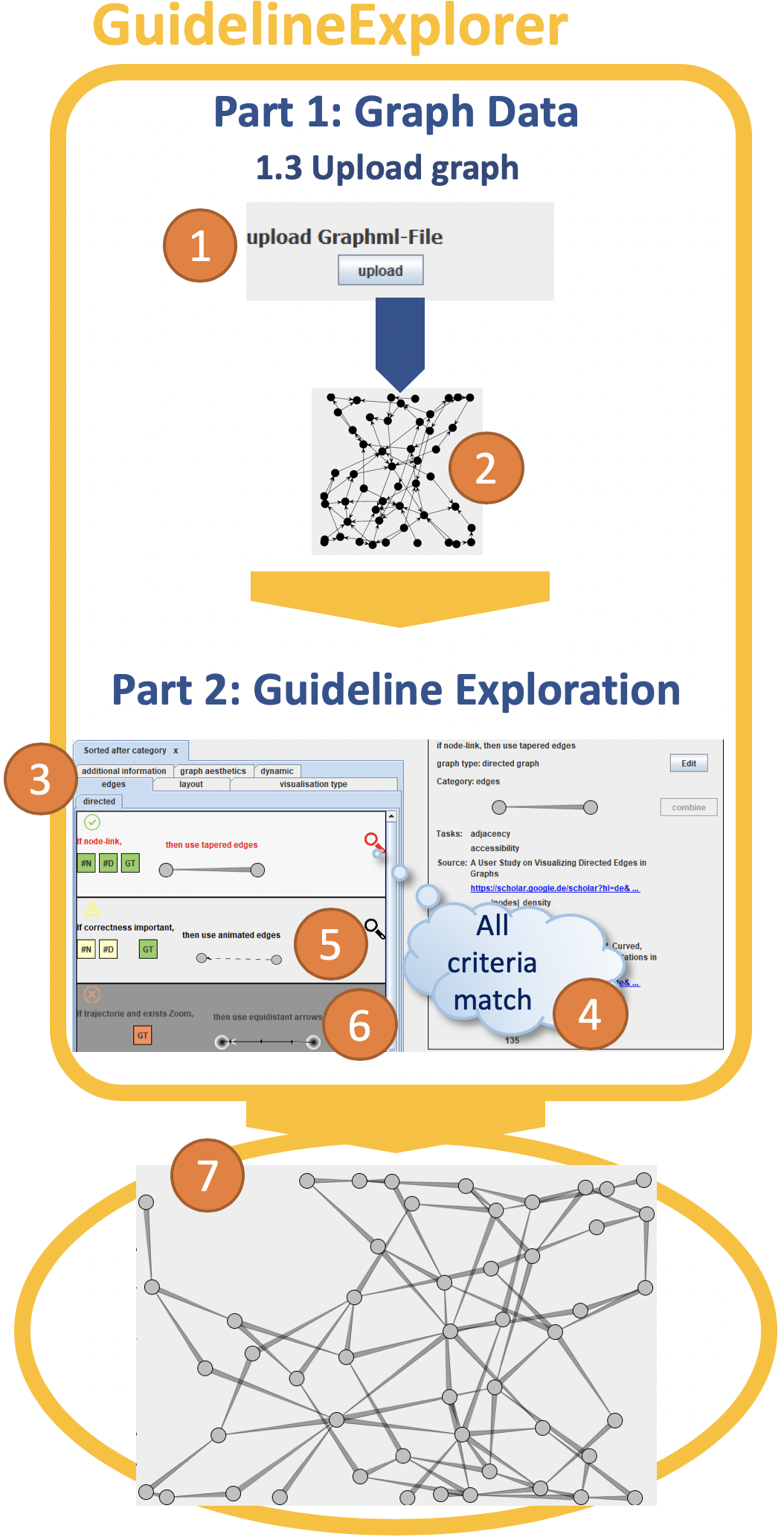}
  \caption[GuidelineExplorer -- use case part 1]{Use case part 1 (based on Figure from \cite{LisaBA}) -- Her first directed graph is equal to the one which Holten et al. \cite{10.1145/1518701.1519054} used in this study (2). After the graph's upload (1), she directly navigates to the directed edge guidelines (3), There, the visualization designer searches for the most suited guideline. She chooses the guideline on using directed edges of Holten et al. \cite{10.1145/1518701.1519054} as all criteria match (4). The remaining guidelines were less suitable (5, 6). The visualization designer is pleased with the visualization result and keeps it (7).}
  \label{fig:pictures_DACH_Chapter3_GuidelineExplorer_UseCase_1}
\end{figure}

Her first directed graph is equal to the one which Holten et al. \cite{10.1145/1518701.1519054} used in this study . It has $50$ nodes and a density of $0.0637$ (cf. Figure~\ref{fig:pictures_DACH_Chapter3_GuidelineExplorer_UseCase_1} -- \texttt{2}). After uploading the graph as a GraphML (cf. Figure~\ref{fig:pictures_DACH_Chapter3_GuidelineExplorer_UseCase_1} -- \texttt{1}), she directly navigates to the directed edge guidelines (cf. Figure~\ref{fig:pictures_DACH_Chapter3_GuidelineExplorer_UseCase_1} -- \texttt{3}). There, the visualization designer searches for the most suited guideline. She chooses the guideline on using directed edges of Holten et al. \cite{10.1145/1518701.1519054} as all criteria match (cf Figure~\ref{fig:pictures_DACH_Chapter3_GuidelineExplorer_UseCase_1} -- \texttt{4}). The other guidelines were less suitable -- cf. e.g., Figure~\ref{fig:pictures_DACH_Chapter3_GuidelineExplorer_UseCase_1} -- \texttt{5}, \texttt{6}. 
As the GuidelineExplorer's visualization result shows (cf. Figure~\ref{fig:pictures_DACH_Chapter3_GuidelineExplorer_UseCase_1} -- \texttt{7}) the first directed graph is well readable visualized with tapered edges. So, the visualization designer keeps the visualization decision recommended by the guideline -- tapered edges.
\FloatBarrier

\begin{figure}[tb]
  \centering
    \includegraphics[width=.8\textwidth]{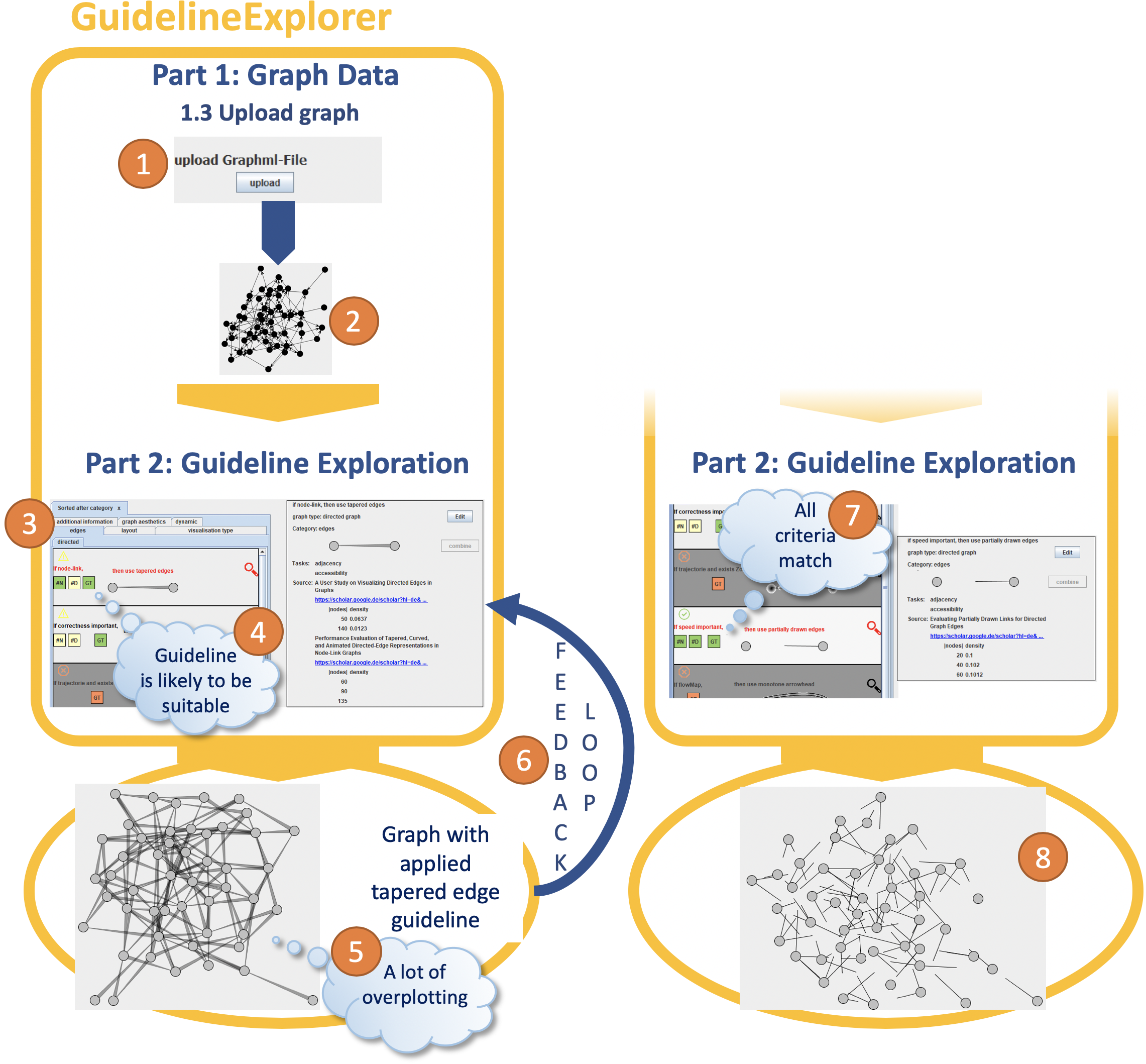}
  \caption[GuidelineExplorer -- use case part 2]{Use case part 2 (based on Figure from \cite{LisaBA}) --  The visualization designer's second graph is denser than her first one (2, cf. Figure~\ref{fig:pictures_DACH_Chapter3_GuidelineExplorer_UseCase_1}). After uploading the graph (1), she navigates again directly to the directed edge guidelines (3). The visualization designer decides for using the tapered edge guideline also for the denser graph as the other two criteria (number of nodes (\#N) and graph type (GT) -- match and the guideline worked so well for her first, but sparser, graph (4). In the final visualization there is a lot of overplotting and visual clutter (5). So, she loops back to the guidelines' overview to inspect the other guidelines (6). She finds that the guideline for partially drawn edges \cite{10.1007/978-3-642-25878-7_22} is even more suitable -- all criteria match (7). Finally, the visualization designer is pleased with the visualization result and keeps it (8).}
  \label{fig:pictures_DACH_Chapter3_GuidelineExplorer_UseCase2}
\end{figure}

The visualization designer's second directed graph which she wants to visualize has $50$ nodes as well, but it is denser (cf. Figure~\ref{fig:pictures_DACH_Chapter3_GuidelineExplorer_UseCase2} -- \texttt{2}). It has a density of $0.1012$. Again, she uploads the graph as a GraphML file (cf. Figure~\ref{fig:pictures_DACH_Chapter3_GuidelineExplorer_UseCase2} -- \texttt{1}). Afterwards, she navigates again directly to the guidelines for directed edges as she is looking for a guideline on how to visualize the direction of directed edges best (cf. Figure~\ref{fig:pictures_DACH_Chapter3_GuidelineExplorer_UseCase2} -- \texttt{3}). The visualization designer, decides for using the tapered edge guideline also for the denser graph as the other two criteria -- number of nodes (\fcolorbox{black}{colorbrewergreen}{\#N}) and graph type \fcolorbox{black}{colorbrewergreen}{GT} -- match and the guideline worked so well for her first, but sparser, graph (cf. Figure~\ref{fig:pictures_DACH_Chapter3_GuidelineExplorer_UseCase2} -- \texttt{4}). However, when looking at the visualization result, cf. Figure~\ref{fig:pictures_DACH_Chapter3_GuidelineExplorer_UseCase2} -- \texttt{5}, she realized that there is a lot of overplotting and visual clutter. Thus, she loops back to the guideline exploration view and searches for even better suited guidelines. After investigating the list view, she quickly finds one. She finds that the guideline recommending partially drawn edges of Burch et al. \cite{10.1007/978-3-642-25878-7_22} matches even better. All three criteria match: number of nodes (\fcolorbox{black}{colorbrewergreen}{\#N}), density (\fcolorbox{black}{colorbrewergreen}{\#D}), and graph type (\fcolorbox{black}{colorbrewergreen}{GT}) (cf. Figure~\ref{fig:pictures_DACH_Chapter3_GuidelineExplorer_UseCase2} -- \texttt{7}). Thus, she applies the guideline. In the visualization result she can see that the partially drawn edges guideline solves the problem of overplotting and visual clutter (cf. Figure~\ref{fig:pictures_DACH_Chapter3_GuidelineExplorer_UseCase2} -- \texttt{8}). So, she keeps this visualization decision for her denser directed graph with $50$ nodes.
\FloatBarrier


\section{Discussion, Conclusion, and Future Work} 
\label{sec:discussion_conclusion_and_future_work}
The contribution within this work is twofold: First, we contribute the taxonomical perspectives on guidelines, which include our foundational perspective -- the visualization decisions necessary to be made for a specific visualization type which are in our case those necessary for a node-link diagram (cf. Section~\ref{sub:basic_taxonomy}) and the perspective by graph type, if-condition type and tasks (cf. Section~\ref{sub:graph_type_perspective},~\ref{sub:if_type_perspective},~\ref{sub:task_perspective}).  Secondly, we contribute the system GuidelineExplorer which implements our taxonomy and an initial set of guidelines.

With the help of our taxonomy it is possible to get an overview of the guidelines and it is easier to find a concrete guideline for a visualization problem. The fact that this is a challenge that needs to be tackled has also been acknowledged by the research community. The workshop ``VisGuides: Workshop on the Creation, Curation, Critique and Conditioning of Principles and Guidelines in Visualization'' \cite{VisGuides_webpage}, which takes place regularly at IEEE VIS, has as current focus the creation, discussion, and implementation of ``[...] [a] framework, or template, or methodology to capture guidelines [...]''.
Such a taxonomy of guidelines like ours can also be transferred to other data resp. visualization types by using the principle that the system uses. We demonstrated this by adding guidelines on adjacency matrices. For example, the principle of our foundational perspective is structuring according to the necessary visualization decisions. In the case of node-link diagrams that we have used in our work, necessary visualization decisions include the visualization of nodes and edges, the layout of the graph, the edges' routing and -- if the graphs have additional information -- the visualization of the additional information. The principle of the other taxonomical perspectives is the type of task, the if-condition or the graph itself.

Our first implementation of the system GuidelineExplorer shows:
\begin{enumerate}
	\item What such a system for exploring the guidelines of a visualization type based on the structure of a taxonomy might look like
	\item The benefits of our proposed taxonomy with its different perspectives and a system like GuidelineExplorer in practice:
	\begin{enumerate}
		\item Transferability of our taxonomy into a system
		\item Simplifying the process of finding and trying out a guideline for a concrete visualization problem. Here, we illustrate this with our use case (cf. Section~\ref{sec:use_case_guidelines})
	\end{enumerate}
\end{enumerate}

Structuring research -- here the guidelines -- according to a taxonomy also helps to identify research white spots. These are research areas which are not yet well covered. We could identify, throughout our search for guideline formulating research, that node guidelines seem to be underrepresented as compared to, for instance, edge guidelines (cf. Sections~\ref{sub:nodes},~\ref{sub:edges}). Further, we could see, by joining the visualization perspective and the task perspective, that directed graphw are less used for research on dynamic graphs as compared to other graph types. Descriptively analyzing the data contained in the task-based taxonomical structures of the guidelines revealed the most popular tasks for guideline research. These are tasks dealing with neighboring nodes, path tracing, and finding the shortest path. A similar analysis for the additional data on the number of nodes we tracked for each guideline shows that for guideline research usually uses graphs with a number of nodes $\leq 80$. Knowledge about the most common tasks and data sizes helps to design future user studies which are comparable to the current body of work. This advantage is also discussed by the state of the art of Yoghourdjian et al. \cite{YOGHOURDJIAN2018264} on the complexity of graphs which are used for user studies. While the state of the art of Yoghourdjian et al. \cite{YOGHOURDJIAN2018264} provides a general overview of graph complexity in user studies, our results are specific for guideline research and no other study types are intermingled.\\
As our research goals were twofold -- taxonomical advances for actionable guidelines and a visual interactive system which allows the exploration and easy application of the very same guidelines -- we did not go into that kind of detail as pure taxonomical papers with a pure theoretical contribution. Consequently, it is necessary to extend our paper pool in the future to i.a. manifest our insights on research white spots. 



  
  \section*{Acknowledgements}

We greatly benefited from the feedback of Prof. Hans-J\"{o}rg Schulz and Prof. G\"{u}nther Wallner. We would like to thank you very much for this.
  
  \bibliographystyle{abbrvurl}
  \bibliography{meine_Diss_references}

\end{document}